# MISALIGNED RING RESONATOR WITH A LENS-LIKE MEDIUM

A.Ya. Bekshaev, V.M. Grimblatov, V.V. Kalugin

*Odessa I.I. Mechnikov National University, Dvorianska 2, Odessa 65082, Ukraine*

The paper presents a theoretical study of the eigenmodes of a misaligned ring multi-mirror laser cavity one or several arms of which are filled with an inhomogeneous medium. We start with posing the general problem of calculation of the radiation characteristics for a ring resonator with arbitrary inhomogeneities of the medium and mirrors, in the paraxial approximation. Then the general relations are specified for the lens-like resonator model in which the real and imaginary parts of the refractive index, as well as the phase and amplitude corrections performed by the mirrors, quadratically depend on the transverse coordinates. Explicit expressions are obtained for the eigen frequency and spatial characteristics of the radiation via the coefficients of the Hermite-Gaussian functions describing the complex amplitude distributions at the mirrors. The main results are formulated in terms of linear relations between positions of the resonator mode axis at the mirrors and the misalignment parameters (small shifts and tilts of the mirrors).

The explicit results display the coefficients of the above linear relations. They are calculated for the 3-mirror cavity with using a specially developed approach employing peculiar conditions to the coefficients' form that follow from very general considerations of two groups:

1) Independence of the resulting frequency shifts on the sequence in which the misalignments are made (this is the base of the "energy method" for the resonator analysis, which is essentially generalized and improved);

2) Geometrical symmetry of the resonator.

The results of calculations can be used to control the radiation characteristics in relation to the resonator misalignments, in particular, to analyze the output radiation stability and sensitivity to the changes in the cavity configuration.

1. Numerous applications of ring lasers require high stability of their radiation [1], which makes it very actual to study how sensitive are their characteristics to diverse perturbations. Among the most important perturbations, practically always present in real situations, are misalignments of the optical structure of the laser cavity (resonator), and many research works deal with the analysis of their influence on laser performance (see, e.g., [2]). This problem is considered rather completely for the case of an "empty" resonator; but in many practical situations, it is necessary to account for the longitudinal and transverse resonator inhomogeneity. It is usually caused by the non-uniform distribution of the refraction index and/or amplification coefficient (gain) within the active medium (including the losses as negative gain) as well as by the important circumstance that the active medium, as a rule, occupies only a part of the resonator perimeter.

It is a difficult task to describe the resonator inhomogeneity in the general form, and usually the study is limited to the case of lens-like cavity filling [2] where the complex wavenumber $k$ of the monochromatic radiation depends on the transverse coordinate $x$ according to the law

$$k(x) = k_0 - \frac{k_2}{2}x^2, \quad k(x) = k'(x) + ik''(x).$$

This model was fruitfully used for the analysis of ring resonators with a transversely inhomogeneous medium arbitrarily situated inside the resonator [3–8] and for the calculation of resonators containing internal diaphragms [9–12], both with ideal alignment and in presence of misalignments [13–16]. Most of the published works concern the three-mirror resonator whose structure frequently occurs in practice.

However, the works mentioned deal with the case of a weakly-inhomogeneous medium where the condition is satisfied

$$\left|\frac{k_2}{k_0}L^2\right| \ll 1$$

where $L$ is the length of the inhomogeneous active medium. But, in agreement with available experimental data and theoretical estimates [17,18], this condition is often violated even in gas-discharge active elements with small amplification (gain). Therefore, in the present work, we study the characteristics of misaligned ring resonators with arbitrary lens-like inhomogeneous medium.

Under such conditions, many visual intuitively clear notions, usually employed to build the models describing the radiation of a misaligned resonator, lose their validity (in particular, in the medium with inhomogeneous amplification one cannot preserve the usual physical meaning of the conception of "ray" [19]). As a result, we need a formalized approach free from familiar intuitive inferences. To this purpose, in Section 2 we present the standard formulation of the problem for calculation of the radiation characteristics in a ring resonator with arbitrary inhomogeneities of the mirrors and the medium. In Sec. 3, the obtained general relations are specified for the lens-like resonator model, and the expressions are derived for the main frequency and spatial characteristics of the radiation via coefficients of the Hermite-Gaussian functions that describe the complex amplitude distribution at the mirrors. The basic result of this Section is the general form of relations between the quantities characterizing the position beam axis with respect to the axial contour of the unperturbed resonator, and the misalignment parameters.

Explicit expressions for the coefficients of these relations are calculated in Sec. 4. Rather bulky calculations can be essentially facilitated due to restrictions for the coefficients' forms that follow from the independence of the resulting shift of the resonator eigenfrequency of the consecutive order in which the mirrors' misalignments are performed (Appendix 1), and from the geometric symmetry of the resonator (Appendix 3). Details of the specific calculations are exposed in Appendix 4.

The results of calculation permit to derive some conclusions on the stability of certain resonator configurations with respect to misalignments and to discuss possible applications of the revealed dependence between the radiation characteristics and the medium inhomogeneity (Sec. 5).

2. Let us consider a plane ring resonator whose axial contour, in the perfectly aligned state, lies within a single plane. The resonator contains arbitrary number of mirrors and arms filled with an inhomogeneous medium. In each arm, we introduce the coordinate frame $X_j$, $Y_j$, $Z_j$ so that axis $Z_j$ coincides with the local position of the unperturbed axial contour, and axis $X_j$ belongs to the axial contour plane and is directed outside (Fig. 1). For the wave running along the axes $Z_j$ the electric and magnetic field strengths in the $j$-th-arm frame can be represented in the form

$$\mathbf{E}_j = \frac{\exp\left[i\Phi(z_j)\right]}{\sqrt{k_0(z_j)}}\left(u_j \mathbf{e}_{Tj} + v_{Ej}\mathbf{e}_{3j}\right), \quad \mathbf{H}_j = \frac{\exp\left[i\Phi(z_j)\right]}{\sqrt{k_0(z_j)}} n_0(z_j)\left(u_j\left[\mathbf{e}_{3j}\times\mathbf{e}_{Tj}\right] + v_{Hj}\mathbf{e}_{3j}\right) \quad (1)$$

($\mathbf{e}_{3j}$, $\mathbf{e}_{Tj}$ are the unit vectors of the longitudinal and transverse directions), where the possible longitudinal variations of the wavenumber is taken into account, $n_0(z_j) = \frac{c}{\omega}k_0(z_j)$ is the refraction index on the $j$-th arm axis, $c$ is the light velocity in vacuum, $\omega$ is the radiation frequency,

$$\Phi(z_j) = \int^{z_j} k_0(z_j)dz_j$$

and functions $u_j(x_j, y_j, z_j)$, $v_j(x_j, y_j, z_j)$ are determined by the quasioptic equations [18,20,21]

$$i\frac{\partial u_j}{\partial z_j} = -\frac{1}{2k_0(z_j)}\nabla_{Tj}^2 u_j + k_0(z_j) V(x_j, y_j, z_j) u_j, \quad (2)$$

$$v_{Ej} = \frac{i}{k_0(z_j)}\left(\mathbf{e}_{Tj}\cdot\nabla_j u_j\right) \quad v_{Hj} = \frac{i}{k_0(z_j)}\left(\mathbf{e}_{3j}\cdot\left[\mathbf{e}_{Tj}\times\nabla_j u_j\right]\right), \quad (3)$$

$$\nabla_{Tj}^2 \equiv \frac{\partial^2}{\partial x_j^2} + \frac{\partial^2}{\partial y_j^2}, \quad \nabla_j \equiv \mathbf{e}_{1j}\frac{\partial}{\partial x_j} + \mathbf{e}_{2j}\frac{\partial}{\partial y_j} + \mathbf{e}_{3j}\frac{\partial}{\partial z_j}.$$

The spatial inhomogeneity of the wavenumber is represented in the form

$$k(x_j, y_j, z_j) = k_0(z_j)\left[1 - V(x_j, y_j, z_j)\right].$$

Functions $u_j$ in adjacent arms are related by the boundary conditions that describe the beam complex amplitude transformation upon reflection at a mirror of arbitrary shape. In the zero-order paraxial approximation [20], the results of Ref. [22] allow us to represent the boundary conditions in the form

$$\rho_\mu = \exp\left[-2ik_{0\mu}\cos\alpha_\mu \Psi_\mu(x,y)\right]\exp\left[i(\Phi_{O\mu} - \Phi_{\Pi\mu})\right]\frac{u_{O\mu}(-x\cos\alpha_\mu + z_\mu\sin\alpha_\mu, y, 0)}{u_{\Pi\mu}(x\cos\alpha_\mu + z_\mu\sin\alpha_\mu, y, 0)} \quad (4)$$

where the subscripts $\Pi$ and O denote the incident and reflected beams, $\alpha_\mu$ is the angle of incidence of the beam axis onto the $\mu$-th mirror in the perfectly aligned resonator, $\rho_\mu$ is the reflection coefficient whose coordinate dependence allows for the spatial inhomogeneity and finite sizes of the mirrors. The subscript $\mu$ denotes that the corresponding quantity characterizes the $\mu$-th mirror or a function of the longitudinal coordinate near the $\mu$-th mirror whose shape in the $XYZ$ frame (Fig. 1) is described by equation

$$z = \Psi_\mu(x,y) = z_\mu + \varphi_\mu(x,y) \quad (5)$$

Conditions of the form (4) can be written for every mirror so the number of such conditions equals to the number of arms the resonator contains. Therefore, the system of equations (2) and (4) is closed and permits to find the field distribution in any cross section of the resonator within a constant multiplier. This multiplier can be further determined with account for the non-linear properties of the resonator medium which in this model are ignored.

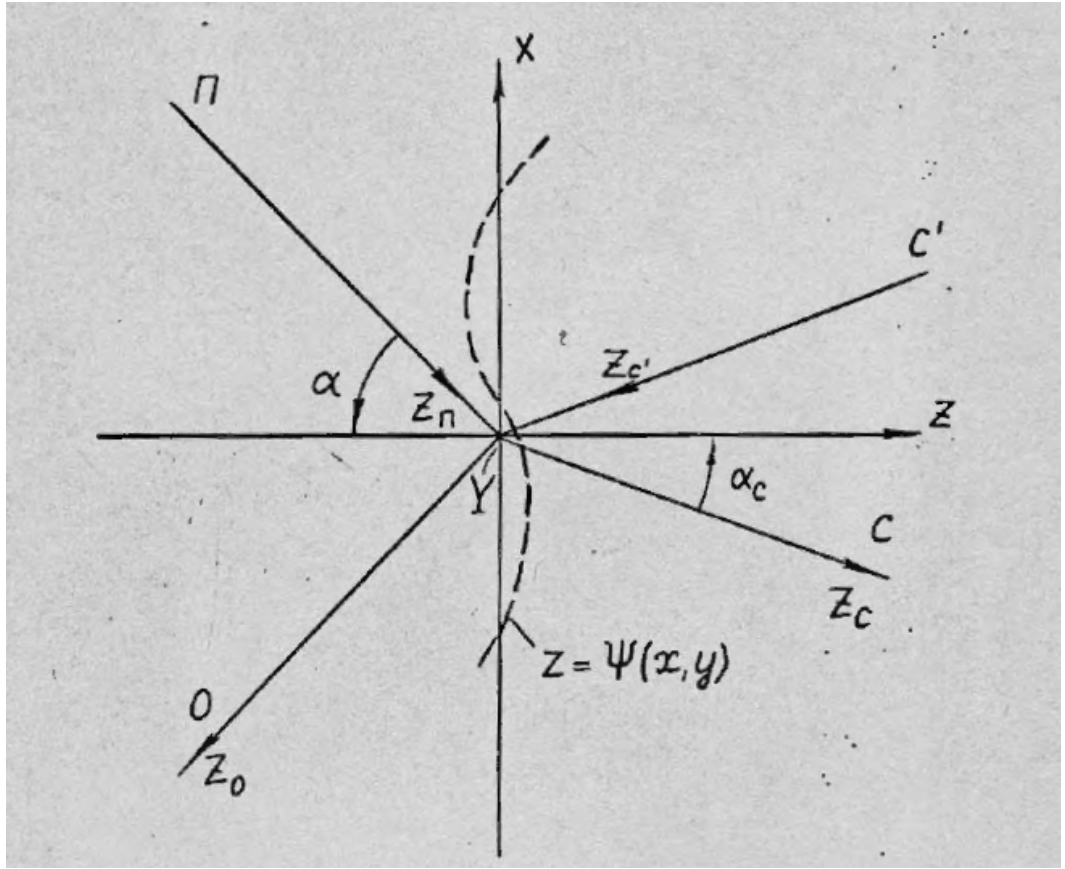

Fig. 1. The coordinate frames associated with the interrelated beams near a mirror. Axes $Z_\Pi$, $Z_O$, $Z_C$ coincide with the nominal axes of the incident, reflected and output beams, axis $Z$ is the nominal normal to the mirror in the nominal point of incidence; axes $Y$ and $Y_\Pi$, $Y_O$, $Y_C$ all coincide and are directed towards the observer, their traces merge in the point $Y$.

Since the longitudinal components of the fields (1) do not enter the boundary conditions (4), they will not occur in further calculations. Nevertheless, expressions (1) and (3) are used for deriving the auxiliary relations in Appendix 1. Now, combining the relations (4) for various mirrors, we obtain the amplitude – phase balance condition

$$\prod_\mu \rho_\mu = \exp\left[-2i\sum_\mu k_{0\mu}\cos\alpha_\mu \Psi_\mu(x,y) - i\Delta\Phi\right]\prod_\mu \frac{u_{O\mu}(-x\cos\alpha_\mu + z_\mu \sin\alpha_\mu, y, 0)}{u_{\Pi\mu}(x\cos\alpha_\mu + z_\mu \sin\alpha_\mu, y, 0)} \quad (6)$$

where

$$\Delta\Phi = \oint k_0(z)\,dz. \quad (7)$$

The quantity $\Delta\Phi$ determines the eigen frequency $\omega$ and the generation threshold $k_0''$ via the relations

$$\omega = \frac{c\,\mathrm{Re}(\Delta\Phi)}{\oint n_0'(z)\,dz}, \quad (8)$$

$$\oint k_0''(z)\,dz = \mathrm{Im}(\Delta\Phi). \quad (9)$$

In further presentation, it will be suitable to express $\Delta\Phi$ via the addition to the value, which this quantity possesses in the empty resonator with the same unperturbed axial contour but composed of plane mirrors:

$$(\Delta\Phi)_g = \Delta\Phi - 2q\pi - i\sum_\mu \ln\rho_\mu, \quad q = 1,2,3,\ldots. \qquad (10)$$

By using the results of Ref. [22] one may, knowing the values of $u_j$ inside the resonator, determine the spatial distribution of the outgoing beams' amplitude and phase immediately after the mirrors, which permits us to find the characteristics of the output radiation and thus to solve completely the problem of the resonator analysis. Note that in the linear approximation, to which the present consideration is restricted, counter-propagating waves of the ring resonator are independent. Therefore, the whole above consideration is equally correct for the wave running along the negative $Z_j$ direction provided that everywhere the signs before $k_0$ and $n_0$ are inverted. Then, the "backward" wave is described in the same frame that was introduced for the "forward" one; the positive direction of the resonator round-trip remains also the same, i.e. the subscript $\Pi$ is already associated with the reflected beam and O – with the incident beam.

3. Now we apply the above relations to the analysis of a misaligned ring resonator. In the lens-like model with separable variables in Eqs. (4) and (5) we should assume

$$\varphi_\mu(x,y) = -\frac{(x-x_\mu)^2 + (y-y_\mu)^2}{2R_\mu} - \beta_\mu^{(x)}(x-x_\mu) - \beta_\mu^{(y)}(y-y_\mu) \qquad (11)$$

where $R_\mu$ is the radius of curvature of the $\mu$-th mirror, and parameters $x_\mu$, $y_\mu$, $\beta_\mu^{(x)}$, $\beta_\mu^{(y)}$ characterize the misalignment. The medium inhomogeneity is given by the function

$$V(x_j, y_j, z_j) = \frac{k_2(z_j)}{2k_0(z_j)}(x_j^2 + y_j^2). \qquad (12)$$

The choice of $V$ in the form (12) means that axis $Z_j$ in the arm containing the active medium is strictly "fastened" to the medium, and the case of the medium misalignment should be considered by introducing the equivalent mirror misalignment. This is always possible if the inhomogeneous medium is present only in a single arm of the resonator. If the resonator contains segments filled with inhomogeneous media whose mutual orientation can alter, one must include into (12) additional terms, linear in $x_j$ and $y_i$, to describe the medium misalignment. In such a case, the resonator analysis can be performed by the same procedure and practically with no additional complications (an analogous problem for the quantum oscillator is considered in [23]).

The wave beam propagating in positive direction of axis $Z_j$ in the $j$-th arm inside the resonator is described by Eqs. (1) in which, according to (12) and (2)

$$u_{jmm'}(x_j, y_j, z_j) = u_{jm}(x_j, z_j) u_{jm'}(y_j, z_j);$$

$$u_{jm}(x_j, z_j) = \exp\left(-\frac{1}{2}a_j^{(x)}x_j^2 + t_j^{(x)}x + f_{mj}^{(x)}\right) H_m\left[\sqrt{p_j^{(x)}}(x_j - \xi_j^{(x)})\right] \qquad (13)$$

where $m$, $m'$ are the mode indices, $H_m$ is the Hermite polynomial. A similar expression takes place for the wave running in the opposite direction.

In each arm of the resonator, the coefficients entering expression (13) obey the equations [24,25]

$$i\frac{da_j}{dz_j} = \frac{a_j^2}{k_0(z_j)} - k_2(z_j), \qquad (14)$$

$$i\frac{dt_j}{dz_j} = \frac{a_j t_j}{k_0(z_j)}, \qquad (15)$$

$$t_j = a_j \xi_j + ik_0(z_j)\frac{d\xi_j}{dz_j}, \qquad (16)$$

$$\frac{d}{dz_j}\left[k_0(z_j)\frac{d\xi_j}{dz_j}\right]+k_2(z_j)\xi_j=0, \tag{17}$$

$$ik_0(z_j)\frac{dp_j}{dz_j}=-2p_j^2+2a_j p_j, \tag{18}$$

$$i\frac{df_{mj}}{dz_j}=\frac{a_j}{2k_0(z_j)}-\frac{t_j^2}{2k_0(z_j)}+m\frac{p_j}{k_0(z_j)}. \tag{19}$$

Boundary conditions relating the parameters of the incident and reflected beams at the $\mu$-th mirror are obtained via the substitution of (13) into (4), and take on the forms (here $\bar{\beta}_\mu^{(\tau)}=\beta_\mu^{(\tau)}-(\tau_\mu/R_\mu)$, $\tau=x,y$)

$$a_{\text{O}\mu}^{(x)}=a_{\Pi\mu}^{(x)}+\frac{2ik_{0\mu}}{R_\mu\cos\alpha_\mu},\quad a_{\text{O}\mu}^{(y)}=a_{\Pi\mu}^{(y)}+\frac{2ik_{0\mu}\cos\alpha_\mu}{R_\mu}; \tag{20}$$

$$t_{\text{O}\mu}^{(x)}=-t_{\Pi\mu}^{(x)}+2ik_{0\mu}\left[a_{\Pi\mu}^{(x)}\frac{z_\mu\sin\alpha_\mu}{ik_{0\mu}}+\bar{\beta}_\mu^{(x)}+\frac{z_\mu}{R_\mu}\tan\alpha_\mu\right],\quad t_{\text{O}\mu}^{(y)}=t_{\Pi\mu}^{(y)}-2ik_{0\mu}\cos\alpha_\mu\cdot\bar{\beta}_\mu^{(y)}; \tag{21}$$

$$\xi_{\text{O}\mu}^{(x)}=-\xi_{\Pi\mu}^{(x)}+2z_\mu\sin\alpha_\mu,\quad \xi_{\text{O}\mu}^{(y)}=\xi_{\Pi\mu}^{(y)}; \tag{22}$$

$$\left(\frac{d\xi_{\text{O}}^{(x)}}{dz_{\text{O}}}\right)_\mu=-\left(\frac{d\xi_{\Pi}^{(x)}}{dz_{\Pi}}\right)_\mu+\frac{2\xi_{\Pi\mu}^{(x)}}{R_\mu\cos\alpha_\mu}+2\bar{\beta}_\mu^{(x)}-\frac{2z_\mu}{R_\mu}\tan\alpha_\mu, \tag{23a}$$

$$\left(\frac{d\xi_{\text{O}}^{(y)}}{dz_{\text{O}}}\right)_\mu=\left(\frac{d\xi_{\Pi}^{(y)}}{dz_{\Pi}}\right)_\mu-2\cos\alpha_\mu\left(\bar{\beta}_\mu^{(y)}+\frac{\xi_{\Pi\mu}^{(y)}}{R_\mu}\right); \tag{23b}$$

$$p_{\text{O}\mu}^{(x)}=p_{\Pi\mu}^{(x)},\quad p_{\text{O}\mu}^{(y)}=p_{\Pi\mu}^{(y)}. \tag{24}$$

For the wave propagating in the opposite direction, the boundary conditions follow from (20) – (24) after the sign reversal before $k_0$, according to what was said in Sec. 2. Formulae (20) – (24) can be also obtained as immediate consequence of the results of Ref. [22].

Solution of the set of equations (14) – (19) with account for the tailoring conditions (20) – (24) supplies the coefficients of expressions (13) for the eigen beams of the resonator. Note that equations for the coefficients characterizing the counter-propagating waves are separable, which confirms that in the lens-like approximation the forward and backward waves are not connected. However, since $p$ and $\xi$ for the counter-propagating beams are determined, as could be easily verified, by the identical equations and boundary conditions, these quantities for both waves coincide.

One can easily see that quantities $a$, $p$ are fully determined by the unperturbed resonator configuration whereas $t$ and $\xi$ are conditioned exclusively by the misalignment and for $z_\mu=\bar{\beta}_\mu=0$, $t=\xi=0$. Like in case of a linear (two-mirror) resonator [24], all the misalignment-depending parameters are determined by the quantity $\xi$, and that is why the misalignment influence on both counter-propagating waves can be studied by considering only one of them, e.g. that is described by Eq. (13).

Beams in the ring resonator obey general regularities inherent in Hermite-Gaussian beams of a resonator with complex lens-likeness [24]. Some corrections in the calculation formulae follow from the different geometry. The most important of them are related to the beam transformation upon its exit from the resonator through a semitransparent mirror. Corresponding formulae similar to (20) – (24) are derived in Ref. [22]. Here we present the relationships describing transformation of the quantity $\xi$ and its derivative with respect to $z$ (subscript C denotes the medium and beam parameters immediately behind the mirror outside the resonator, parameters of the output beam correspond to its representation in the form (13) with $j = C$ in the coordinate frame $X_C\ Y_C\ Z_C$, see Fig. 1):

$$\frac{\xi_{C\mu}^{(x)}}{\cos\alpha_{C\mu}} - \frac{\xi_{\Pi\mu}^{(x)}}{\cos\alpha_\mu} = z_\mu \frac{\Delta_\mu \tan\alpha_{C\mu}}{k_{0\mu}\cos\alpha_\mu}, \quad \xi_{C\mu}^{(y)} = \xi_{\Pi\mu}^{(y)}, \tag{25}$$

$$k_{0C}\left(\frac{d\xi_C^{(x)}}{dz_C}\right)_\mu \cos\alpha_{C\mu} - k_{0\mu}\left(\frac{d\xi_\Pi^{(x)}}{dz_\Pi}\right)_\mu \cos\alpha_\mu = \Delta_\mu\left(\frac{z_\mu}{R_\mu}\tan\alpha_\mu - \overline{\beta}_\mu^{(x)} - \frac{\xi_{\Pi\mu}^{(x)}}{R_\mu \cos\alpha_\mu}\right),$$

$$k_{0C}\left(\frac{d\xi_C^{(y)}}{dz_C}\right)_\mu - k_{0\mu}\left(\frac{d\xi_\Pi^{(y)}}{dz_\Pi}\right)_\mu = -\Delta_\mu\left(\overline{\beta}_\mu^{(y)} + \frac{\xi_{\Pi\mu}^{(y)}}{R_\mu}\right), \tag{26}$$

$$\Delta_\mu = k_{0\mu}\cos\alpha_\mu - k_{0C}\cos\alpha_{C\mu}.$$

Equations for determining the parameters of the output beam formed by the wave propagating in the opposite direction can be obtained from (25) by replacements $\xi_\Pi \to \xi_O$, $z_\Pi \to z_O$, $\alpha_\mu \to \pi - \alpha_\mu$ and $\alpha_{C\mu} \to \pi - \alpha_{C\mu}$ with simultaneous sign reversal before $k_0$. The parameters of the output beam describe it in the frame $X_{C'}\ Y_{C'}\ Z_{C'}$ (Fig. 1) where it propagates in the negative direction of axis $Z_{C'}$.

As is known, the fundamental-mode-beam axis experiences refraction upon exit from the resonator that does not obey the Snell's law [24]. It can be described by relations (25) and (26) provided that there are made substitutions

$$\xi_C^{(\tau)} \to \tau_{C0}, \quad \xi_\Pi^{(\tau)} \to \tau_{\Pi 0}, \quad \frac{d\xi_C^{(\tau)}}{dz_C} \to \frac{d\tau_{C0}}{dz_C} = \frac{\operatorname{Im}\left(t_C^{(\tau)} a_C^{(\tau)*}\right)}{k_{0C}\operatorname{Re} a_C^{(\tau)}}, \quad \frac{d\xi_\Pi^{(\tau)}}{dz_\Pi} \to \frac{d\tau_{\Pi 0}}{dz_\Pi} = \frac{\operatorname{Im}\left(t_\Pi^{(\tau)} a_\Pi^{(\tau)*}\right)}{k_{0\mu}\operatorname{Re} a_\Pi^{(\tau)}} \tag{27}$$

where $\tau$ stands for $x$, $y$, $\tau_{0C}$ and $\tau_{0\Pi}$ are the transverse coordinates of the incident and outgoing Gaussian beams' axes immediately after and before the mirror in the coordinate frames $X_C\ Y_C\ Z_C$ and $X_\Pi\ Y_\Pi\ Z_\Pi$, correspondingly. Note that when the amplification or loss inhomogeneity is absent (Im$k_2 = 0$), $\tau_{\Pi 0} = \xi_\Pi^{(\tau)}$ and $\tau_{C0} = \xi_C^{(\tau)}$, i.e. the quantity $\xi$ directly represents the position of the beam axis.

Except the specific forms of the boundary conditions, the problems of calculation of the beam characteristics in planes $XZ$ and $YZ$ are absolutely identical. Therefore, further we will consider only the constituents depending on $x$, and suppose that all the misalignment components lie in plane $XZ$; accordingly, the superscript ($x$) will be omitted.

The linear character of Eqs. (15), (17) dictates the linear correspondence between the output beams' characteristics and the misalignment parameters. The most important expressions of this correspondence are the formulae for the values of $\xi$ and $(d\xi/dz)$ at the mirrors which are analogues of the known results for the linear resonator [24–26]

$$\xi_{\Pi\mu} = \sum_\nu \left(a_{\mu\nu}\overline{\beta}_\nu + c_{\mu\nu}z_\nu\right), \tag{28}$$

$$\left(\frac{d\xi_\Pi}{dz_\Pi}\right)_\mu = \sum_\nu \left(a'_{\mu\nu}\overline{\beta}_\nu + c'_{\mu\nu}z_\nu\right), \tag{29}$$

$$\xi_{O\mu} = \sum_\nu \left(b_{\mu\nu}\overline{\beta}_\nu + d_{\mu\nu}z_\nu\right), \tag{30}$$

$$\left(\frac{d\xi_O}{dz_O}\right)_\mu = \sum_\nu \left(b'_{\mu\nu}\overline{\beta}_\nu + d'_{\mu\nu}z_\nu\right) \tag{31}$$

where $a_{\mu\nu}$, $b_{\mu\nu}$, $c_{\mu\nu}$, $d_{\mu\nu}$ and $a'_{\mu\nu}$, $b'_{\mu\nu}$, $c'_{\mu\nu}$, $d'_{\mu\nu}$ are the elements of certain square matrices $(a)=\{a_{\mu\nu}\}$, $(b)=\{b_{\mu\nu}\}$, etc., with dimension equal to the number of mirrors (importantly: matrix elements $a_{\mu\nu}$ and $a'_{\mu\nu}$ of (28), (29) (always with double numerical subscripts) should not be confused with the beam parameters $a_j$, $a^{(x)}_{\Pi\mu}$, etc. of (13) – (21), (27)). We emphasize that the matrices $(a)$, $(c)$, $(a')$, $(c')$ describe the behavior of the beam propagating in the positive direction while $(b)$, $(d)$, $(b')$, $(d')$ describe the beam propagating in the negative direction of the resonator round-trip.

For calculation of the frequency characteristics one should find the correction (10). In the two-dimensional case, for the beam (13), by using the amplitude-phase balance condition (6) and the boundary conditions (20), (21), we find

$$i(\Delta\Phi)_g = -2i\sum_\mu k_{0\mu}\cos\alpha_\mu (z_\mu + \gamma_\mu) + \frac{1}{2}\sum_\mu z_\mu^2 \sin^2\alpha_\mu (a_{\Pi\mu} - a_{O\mu})$$
$$+ \sum_\mu (t_{O\mu} - t_{\Pi\mu}) z_\mu \sin\alpha_\mu - \sum_j \Delta f_j \qquad (32)$$

where

$$\gamma_\mu = -\frac{x_\mu^2}{2R_\mu} + \beta_\mu x_\mu$$

and $\Delta f_j$ is the increment of the quantity $f_j$ on the length of the $j$-th arm that, in agreement to (19), equals to

$$\Delta f_j = \frac{1}{i}\int\left(\frac{a_j}{2} + p_j\right)\frac{dz_j}{k_0(z_j)} - \frac{1}{2i}\int\frac{t_j^2}{k_0(z_j)}dz_j.$$

Now we can divide expression (32) into two parts one of which,

$$(\Delta\Phi)_1 = \sum_j \int\left(\frac{a_j}{2} + p_j\right)\frac{dz_j}{k_0(z_j)}$$

does not depend on the misalignment. Consider the second part. With taking into account the equality

$$-\frac{1}{i}\frac{t_j^2}{2k_0(z_j)} = -\frac{1}{2}\frac{d}{dz_j}(a_j\xi_j^2) - \frac{i}{2}\frac{d}{dz_j}\left[k_0(z_j)\xi_j\frac{d\xi_j}{dz_j}\right]$$

as well as relations (16) and (21) – (23), it can be written in the form

$$(\Delta\Phi)_2 = \sum_\mu k_{0\mu}\left[-2z_\mu\cos\alpha_\mu - \frac{z_\mu^2\sin^2\alpha_\mu}{R_\mu\cos\alpha_\mu} - 2\gamma_\mu\cos\alpha_\mu + \right.$$
$$\left. + \frac{\xi_{\Pi\mu}z_\mu}{R_\mu}\tan\alpha_\mu - \left(\frac{d\xi_\Pi}{dz_\Pi}\right)_\mu z_\mu\sin\alpha_\mu + \xi_{\Pi\mu}\bar{\beta}_\mu\right]. \qquad (33)$$

With employment of (28) and (29), this second part can be represented as a linear-quadratic form

$$(\Delta\Phi)_2 = -2\sum_\mu k_{0\mu}(z_\mu + \gamma_\mu)\cos\alpha_\mu + \sum_{\mu,\nu}\left\{\left[k_{0\mu}\sin\alpha_\mu\left(\frac{a_{\mu\nu}}{R_\mu\cos\alpha_\mu} - a'_{\mu\nu}\right) + k_{0\nu}c_{\nu\mu}\right]\bar{\beta}_\nu z_\mu \right.$$
$$\left. + k_{0\mu}\sin\alpha_\mu\left(\frac{c_{\mu\nu}}{R_\mu\cos\alpha_\mu} + c'_{\mu\nu} - \frac{\tan\alpha_\mu}{R_\mu}\delta_{\mu\nu}\right)z_\mu z_\nu + k_{0\mu}a_{\mu\nu}\bar{\beta}_\mu\bar{\beta}_\nu\right\}, \qquad (34)$$

$\delta_{\mu\nu}$ is the Kronecker symbol.

During calculation of $\Delta\Phi$ for the backward wave, one would start from the relation similar to (6) in which the sign of the exponent is reverted in both parts. Accordingly, the result for $(\Delta\Phi)_2$ of the backward wave coincides with (33). Therefore, knowledge of the matrices (28), (29) permits to completely describe the influence of misalignments on the radiation frequency characteristics. The equations (33) and (34) relate the sensitivity of the resonator eigen frequencies to misalignments with the behavior of the beams' spatial characteristics.

The results of this section demonstrate that the misalignment does not cause the frequency non-reciprocity of the counter-propagating waves. This is because our boundary conditions (4) are valid in the zero paraxial order. Known calculations leading to the contrary conclusion [14,16] are based on the boundary conditions of the type (4) which include some terms of higher paraxial orders [20,21] due to which, in Eqs. (20), an additional misalignment-induced term appears. Letting aside the detailed discussion of the mentioned calculations, we note that the consistent allowance for such terms would require to consider the longitudinal field components, to take into account the non-Gaussian form of the incident and reflected beams, forbid separation of the transverse variables, and would make the whole procedure of the resonator analysis much more complicated [22].

4. Now we employ the developed general scheme for calculation of misalignment characteristics of a set of three-mirror ring resonators with lens-like active medium. Our task is to determine the matrix elements (28) – (31) whose knowledge, in agreement with (34), permits to find the eigen frequency of the misaligned resonator, and in case of a "pure" refraction inhomogeneity (zero imaginary part of $k_2$), also the behavior of the output beams' axes (via (25) and (26)).

Let us consider the resonator formed by one spherical (radius of curvature $R$) and two plane mirrors placed in the vertices of an equilateral triangle with the side length $l$ (Fig. 2b). Let $k_0$ = const along the whole resonator perimeter. Then the boundary conditions (22), (23), with account for the equalities $\sin\alpha_\mu = 0.5$, $\cos\alpha_\mu = \sqrt{3}/2$, will read

$$\xi_{O\mu} = -\xi_{\Pi\mu} + z_\mu, \quad \mu = 1, 2, 3;$$

$$\left(\frac{d\xi_O}{dz_O}\right)_1 = -\left(\frac{d\xi_\Pi}{dz_\Pi}\right)_1 + \frac{4\xi_{\Pi 1}}{R\sqrt{3}} + 2\bar{\beta}_1 - \frac{2z_1}{R\sqrt{3}}; \quad (35)$$

$$\left(\frac{d\xi_O}{dz_O}\right)_\mu = -\left(\frac{d\xi_\Pi}{dz_\Pi}\right)_\mu + 2\bar{\beta}_\mu, \quad \mu = 2, 3.$$

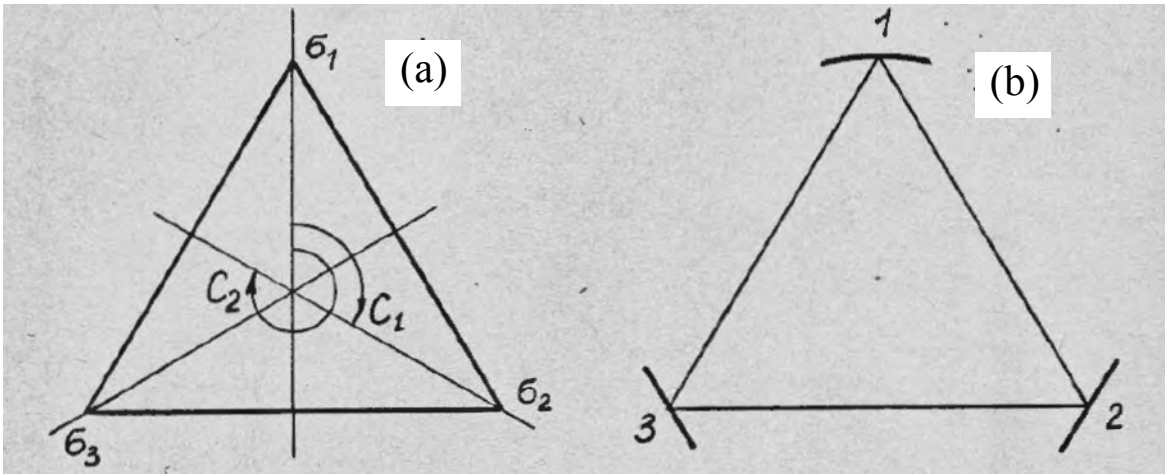

Fig. 2. (a) Symmetry elements of the equilateral triangle; (b) the three-mirror resonator geometry.

The calculations are simplified by the use of equations (A6) from Appendix 1 which in the present case give

$$a_{\mu\nu} = a_{\nu\mu};\tag{36}$$

$$(c) = \frac{1}{2}\begin{pmatrix} 2 - a'_{11} + \frac{2a_{11}}{R\sqrt{3}} & -a'_{21} & -a'_{31} \\ -a'_{12} + \frac{2a_{12}}{R\sqrt{3}} & 2 - a'_{22} & -a'_{32} \\ -a'_{13} + \frac{2a_{13}}{R\sqrt{3}} & -a'_{23} & 2 - a'_{33} \end{pmatrix}\tag{37}$$

(due to which the elements $c_{\mu\nu}$ can be found once $a_{\mu\nu}$ and $a'_{\nu\mu}$ are known);

$$c'_{12} = c'_{21} = \frac{2c_{12}}{R\sqrt{3}}, \quad c'_{13} = c'_{31} = \frac{2c_{13}}{R\sqrt{3}}, \quad c'_{23} = c'_{32}.\tag{38}$$

Similar relation connect the corresponding elements of the matrices $(b)$, $(b')$, $(d)$ and $(d')$ and can be obtained from (36) – (38) upon replacements $a_{\mu\nu} \to -b_{\mu\nu}$, $a'_{\mu\nu} \to b'_{\mu\nu}$, $c_{\mu\nu} \to d_{\mu\nu}$, $c'_{\mu\nu} \to -d'_{\mu\nu}$. Besides, the boundary conditions (35) yield

$$a_{\mu\nu} = -b_{\mu\nu}, \quad d_{\mu\nu} = -c_{\mu\nu} + \delta_{\mu\nu} \quad (\mu, \nu = 1, 2, 3);\tag{39}$$

$$d'_{\mu\nu} = -c'_{\mu\nu}, \quad a'_{\mu\nu} = -b'_{\mu\nu} + 2\delta_{\mu\nu} \quad (\mu = 2, 3; \ \nu = 1, 2, 3);\tag{40}$$

$$\begin{cases} d'_{1\nu} = -c'_{1\nu} + \dfrac{4c_{1\nu}}{R\sqrt{3}} - \dfrac{2}{R\sqrt{3}}\delta_{1\nu}, \\ b'_{1\nu} = -a'_{1\nu} + \dfrac{4a_{1\nu}}{R\sqrt{3}} + 2\delta_{1\nu}, \end{cases} \quad \nu = 1, 2, 3.\tag{41}$$

Equations (36) – (38) reduce the number of independently determined quantities characterizing the forward beam (propagating in the positive direction) from 36 to 21, and Eqs. (39) – (41) give way to application of these quantities for the calculation of matrices describing the backward beam. Therefore, the number of independently determined parameters reduces more than three times. Further simplifications are possible only upon consideration of specific configurations of the filled resonator. In particular, the explicit allowance for the symmetry requirements is rather advantageous (see Appendix 3).

The corresponding calculations described in Appendix 4 supply the following explicit expressions for the matrices (28) – (31).

(a) Empty resonator (all three arms are filled with a homogeneous medium).

$$(a) = \begin{pmatrix} a_{11} & a_{12} & a_{12} \\ a_{12} & a_{22} & a_{23} \\ a_{12} & a_{23} & a_{22} \end{pmatrix}, \quad (b) = -(a);$$

$$(a') = \begin{pmatrix} a'_{11} & a'_{12} & 2 + a'_{12} \\ a'_{11} & 2 + a'_{12} & a'_{12} \\ -a'_{11} & -a'_{12} & -a'_{12} \end{pmatrix}, \quad (b') = \begin{pmatrix} a'_{11} & 2 + a'_{12} & a'_{12} \\ -a'_{11} & -a'_{12} & -a'_{12} \\ a'_{11} & a'_{12} & 2 + a'_{12} \end{pmatrix};$$

$$(c) = \frac{1}{2}\begin{pmatrix} 1 & a'_{11} & a'_{11} \\ 1 & -a'_{12} & a'_{12} \\ -1 & -a'_{12} & 2 + a'_{12} \end{pmatrix}, \quad (d) = \frac{1}{2}\begin{pmatrix} 1 & a'_{11} & -a'_{11} \\ -1 & 2 + a'_{12} & -a'_{12} \\ 1 & a'_{12} & -a'_{12} \end{pmatrix};$$

$$(c') = -\frac{a'_{11}}{R\sqrt{3}}\begin{pmatrix} 0 & 1 & -1 \\ 0 & 1 & -1 \\ 0 & -1 & 1 \end{pmatrix}, \quad (d') = -\frac{a'_{11}}{R\sqrt{3}}\begin{pmatrix} 0 & 1 & -1 \\ 0 & -1 & 1 \\ 0 & 1 & -1 \end{pmatrix}.$$

Here the elements of matrices $(a)$ and $(a')$ are related by equations

$$a'_{11} = 1 + \frac{2a_{11}}{R\sqrt{3}}, \quad a'_{12} = -1 + \frac{2a_{12}}{R\sqrt{3}}$$

and there are only four independent elements, and they equal to

$$a_{11} = -\frac{3l}{D_0}, \quad a_{12} = \frac{l}{D_0}, \quad a_{22} = \frac{l}{D_0}\left(-3 + \frac{8l}{R\sqrt{3}}\right), \quad a_{23} = \frac{l}{D_0}\left(1 - \frac{4l}{R\sqrt{3}}\right)$$

where

$$D_0 = -2 + \frac{6l}{R\sqrt{3}}. \tag{42}$$

(b) Resonator with the lens-like active arm placed between the mirrors 2 and 3 (Fig. 2b); there is a plane of symmetry (the triangle attitude).

$$(a) = \begin{pmatrix} a_{11} & a_{12} & a_{12} \\ a_{12} & a_{22} & a_{23} \\ a_{12} & a_{23} & a_{22} \end{pmatrix}, \quad (b) = -(a);$$

$$(a') = \begin{pmatrix} a'_{11} & a'_{12} & a'_{13} \\ a'_{11} & a'_{13} & a'_{12} \\ -a'_{11} & -a'_{12} & 2 - a'_{13} \end{pmatrix}, \quad (b') = \begin{pmatrix} a'_{11} & a'_{13} & a'_{12} \\ -a'_{11} & 2 - a'_{13} & -a'_{12} \\ a'_{11} & a'_{12} & a'_{13} \end{pmatrix};$$

where $a'_{11} = 1 + \frac{2a_{11}}{R\sqrt{3}}$, $a'_{13} + a'_{12} = 1 + \frac{2a_{12}}{R\sqrt{3}}$;

$$(c) = \begin{pmatrix} \frac{1}{2} & c_{12} & -c_{12} \\ c_{21} & c_{22} & c_{23} \\ -c_{21} & -c_{23} & 1 - c_{22} \end{pmatrix}, \quad (d) = \begin{pmatrix} \frac{1}{2} & -c_{12} & c_{12} \\ -c_{21} & 1 - c_{22} & -c_{23} \\ c_{21} & c_{23} & c_{22} \end{pmatrix}$$

where $c_{12} = -\frac{a'_{11}}{2}$, $c_{22} = 1 - \frac{a'_{13}}{2}$, $c_{23} = \frac{a'_{12}}{2}$, $c_{21} = -\frac{a'_{12}}{2} + \frac{a_{12}}{R\sqrt{3}}$;

$$(c') = \begin{pmatrix} c'_{11} & c'_{12} & c'_{13} \\ -c'_{11} & -c'_{13} & -c'_{12} \\ -c'_{11} & -c'_{12} & -c'_{13} \end{pmatrix}, \quad (d') = \begin{pmatrix} -c'_{11} & -c'_{13} & -c'_{12} \\ c'_{11} & c'_{13} & c'_{12} \\ c'_{11} & c'_{12} & c'_{13} \end{pmatrix}$$

where $c'_{12} = -c'_{11} - \frac{a'_{11}}{R\sqrt{3}}$, $c'_{13} = -c'_{11} + \frac{a'_{11}}{R\sqrt{3}}$.

Only six quantities should be determined independently. If the active medium is longitudinally homogeneous ($k_2(z)$ = const), it can be characterized by the parameter $\eta = \sqrt{k_2/k_0}$ [24], and these six quantities are

$$a_{11} = -\frac{l}{D_c}\left(\frac{\sin\eta l}{\eta l} + 2\cos\eta l - \eta l \sin\eta l\right), \quad a_{12} = -\frac{l}{D_c}\left(1 - \cos\eta l - \frac{\sin\eta l}{\eta l}\right),$$

$$a_{22} = -\frac{l}{D_c}\left[2\cos\eta l + \frac{\sin\eta l}{\eta l} - \frac{4l}{R\sqrt{3}}\left(\cos\eta l + \frac{\sin\eta l}{\eta l}\right)\right], \quad a_{23} = -\frac{l}{D_c}\left(\frac{\sin\eta l}{\eta l} - 2 + \frac{4l}{R\sqrt{3}}\right),$$

$$a'_{12} = \frac{1}{D_c}\left(1 + \cos\eta l - \frac{4l}{R\sqrt{3}}\right), \quad c'_{11} = -\frac{1}{D_c}\left(\frac{1 - \cos\eta l + \eta l \sin\eta l}{R\sqrt{3}} - \frac{\eta}{2}\sin\eta l\right)$$

where

$$D_c = -1 - \cos\eta l + \eta l \sin\eta l + \frac{2l}{R\sqrt{3}}\left(\frac{\sin\eta l}{\eta l} - \eta l \sin\eta l + 2\cos\eta l\right). \tag{43}$$

(c) Resonator with the lens-like arm between the spherical mirror and the plane mirror 2 (Fig. 2b), possessing no geometrical symmetry.

$$(a) = \begin{pmatrix} a_{11} & a_{12} & a_{13} \\ a_{12} & a_{22} & a_{23} \\ a_{13} & a_{23} & a_{33} \end{pmatrix}, \quad (b) = -(a);$$

$$(a') = \begin{pmatrix} a'_{11} & a'_{12} & 2 - a'_{33} \\ a'_{11} & 2 + a'_{12} & -a'_{33} \\ -a'_{11} & -a'_{12} & a'_{33} \end{pmatrix}, \quad (b') = \begin{pmatrix} b'_{11} & b'_{12} & b'_{13} \\ -a'_{11} & -a'_{12} & a'_{33} \\ a'_{11} & a'_{12} & 2 - a'_{33} \end{pmatrix}$$

where $b'_{11} = 2 - a'_{11} + \frac{4a_{11}}{R\sqrt{3}}$, $b'_{12} = -a'_{12} + \frac{4a_{12}}{R\sqrt{3}}$, $b'_{13} = a'_{33} - 2 + \frac{4a_{13}}{R\sqrt{3}}$;

$$(c) = \begin{pmatrix} c_{11} & c_{12} & -c_{12} \\ c_{21} & c_{22} & -c_{22} \\ c_{31} & c_{32} & 1 - c_{32} \end{pmatrix}, \quad (d) = \begin{pmatrix} 1 - c_{11} & -c_{12} & c_{12} \\ -c_{21} & 1 - c_{22} & c_{22} \\ -c_{31} & -c_{32} & c_{32} \end{pmatrix}$$

where

$$c_{11} = 1 - \frac{a'_{11}}{2} + \frac{a_{11}}{R\sqrt{3}}, \quad c_{12} = -\frac{a'_{11}}{2}, \quad c_{21} = -\frac{a'_{12}}{2} + \frac{a_{12}}{R\sqrt{3}}, \quad c_{22} = -\frac{a'_{12}}{2},$$

$$c_{31} = -1 + \frac{a'_{33}}{2} + \frac{a_{13}}{R\sqrt{3}}, \quad c_{32} = \frac{a'_{33}}{2};$$

$$(c') = \begin{pmatrix} c'_{11} & c'_{12} & -c'_{12} \\ c'_{11} & c'_{12} & -c'_{12} \\ -c'_{11} & -c'_{12} & c'_{12} \end{pmatrix}, \quad (d') = \begin{pmatrix} d'_{11} & d'_{12} & -d'_{12} \\ -c'_{11} & -c'_{12} & c'_{12} \\ c'_{11} & c'_{12} & -c'_{12} \end{pmatrix}$$

where $c'_{12} = c'_{11} - \frac{a'_{11}}{R\sqrt{3}}$, $d'_{12} = -c'_{11} - \frac{a'_{11}}{R\sqrt{3}}$.

Here the eleven matrix elements are independent:

$$a_{11} = -\frac{l}{D_n}\left(2\cos\eta l + \frac{\sin\eta l}{\eta l}\right), \quad a_{12} = \frac{l}{D_n}\left(2 - \frac{\sin\eta l}{\eta l}\right), \quad a_{13} = \frac{l}{D_n}\left(\cos\eta l - 1 + \frac{\sin\eta l}{\eta l}\right),$$

$$a_{23} = -\frac{l}{D_n}\left(1 - \cos\eta l - \frac{\sin\eta l}{\eta l} + \frac{4\sin\eta l}{\eta R\sqrt{3}}\right), \quad a_{22} = -\frac{l}{D_n}\left(2\cos\eta l + \frac{\sin\eta l}{\eta l} - \frac{8\sin\eta l}{\eta R\sqrt{3}}\right),$$

$$a_{33} = -\frac{l}{D_n}\left[2\cos\eta l + \frac{\sin\eta l}{\eta l} - \eta l\sin\eta l - \frac{4l}{R\sqrt{3}}\left(\cos\eta l + \frac{\sin\eta l}{\eta l}\right)\right],$$

$$a'_{11} = -\frac{1}{D_n}(1 + \cos\eta l), \quad a'_{12} = \frac{1}{D_n}\left(1 + \cos\eta l - \frac{4\sin\eta l}{\eta R\sqrt{3}}\right),$$

$$a'_{33} = \frac{1}{D_n}\left(\eta l \sin \eta l - 1 - \cos \eta l + \frac{4l \sin \eta l}{R\sqrt{3}}\right),$$

$$c'_{11} = -\frac{1}{D_n}\left(\frac{1-\cos\eta l}{R\sqrt{3}} - \frac{\eta}{2}\sin\eta l\right), \quad d'_{11} = \frac{1}{D_n}\left(\frac{2\eta l \sin \eta l + 1 - \cos \eta l}{R\sqrt{3}} - \frac{\eta}{2}\sin\eta l\right)$$

where

$$D_n = \eta l \sin \eta l - 1 - \cos \eta l + \frac{2l}{R\sqrt{3}}\left(2\cos \eta l + \frac{\sin \eta l}{\eta l}\right). \tag{44}$$

5. Results of Sec. 4 immediately characterize the sensitivity of the output beams' axes to the resonator misalignments. They enable to calculate the axial contour perturbation for arbitrary mirror displacements and any ideal configuration satisfying the conditions accepted (Fig. 2b), except such combinations of $R$, $l$ and $\eta$ for which the quantities $D_0$, $D_c$ and $D_n$ of Eqs. (42) – (44) vanish. These configurations correspond to zero axial contour robustness and are unstable with respect to misalignments. In view of the practical importance of knowing such configurations, in Fig. 3 the curves are shown that represent the geometric loci of points with zero axial contour stability for the three-mirror resonators with symmetric and asymmetric positions of the active medium (cases (b) and (c) of the previous section), in coordinates $R/l$ and $\bar{k}_2 = (\eta l)^2$. The range of $\bar{k}_2$ values embraces the most physically available inhomogeneities of the refraction index.

For the symmetric resonator (case (b) of Sec. 4) instability occurs at the following set of conditions:

$$\eta l = \cot\frac{\eta l}{2}, \quad R \text{ arbitrary};$$

$$\frac{R}{l} = \frac{2}{\sqrt{3}}\left(1 + \frac{1}{\eta l}\tan\frac{\eta l}{2}\right). \tag{45}$$

The first condition is realized at $\eta l = 1.307$, and in Fig. 3 the corresponding curve is the vertical straight line. Note that fulfillment of conditions (45) does not mean that all the matrix elements tend to infinity. Sometimes a sort of relative stability is observed: for example, at first condition (45) the beam is stable with respect to tilts of the spherical mirror but unstable with respect to all other kinds of misalignment. This phenomenon has no analogue in case of linear resonators where under conditions of zero axial contour robustness arbitrarily small misalignments lead to the infinite displacements of the beams' axes [21].

The instability condition for the asymmetric resonator (case (c) of Sec. 4) takes the form

$$\frac{R}{l} = \frac{2}{\sqrt{3}} \frac{\frac{\sin \eta l}{\eta l} + 2\cos \eta l}{1 + \cos \eta l - \eta l \sin \eta l}. \tag{46}$$

It is "absolute" in the meaning that fulfillment of (46) makes singular all the matrix elements (28) – (31).

For illustration of the matrix elements dependence on the "optical strength" $\bar{k}_2$ of the lens-like medium, in Fig. 4 the curves for $a'_{33}$ are presented. In general, their dependence is similar to what was found for linear resonators [26] although discrepancies are seen in some details.

First of all notice that the discontinuities corresponding to zero robustness of the axial contour occur at $k_2 > 0$ (rather than at $k_2 < 0$ as take place for linear resonators); at $k_2 < 0$, on the contrary, the absolute stability to misalignments is observed: for the condition $R = 3.46l$ accepted in Fig. 4, in both cases, with growing $|k_2|$, $a'_{33}$ continuously approaches the asymptotic values 1 and 2 without any singularity.

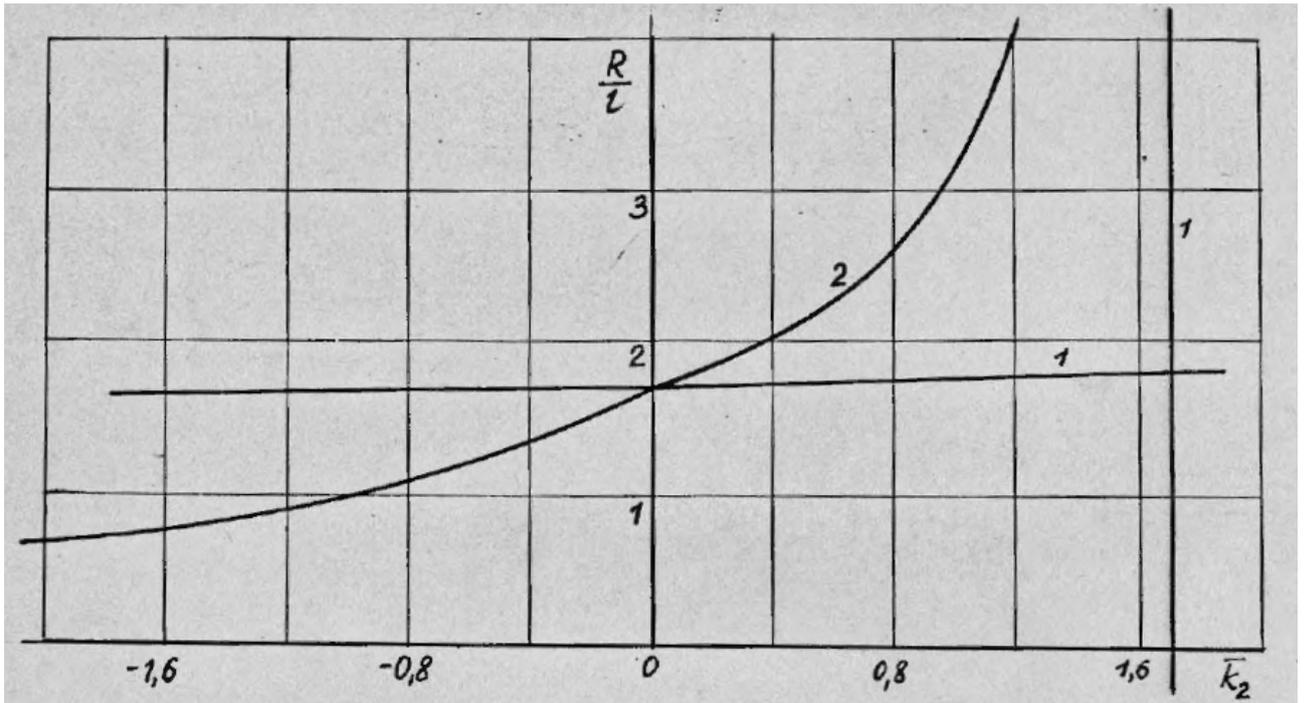

Fig. 3. Curves of zero axial contour robustness for three-mirror resonators with different active element positions (cf. Fig. 2): (1) between the mirrors 2 and 3, (2) between the mirrors 3 and 1.

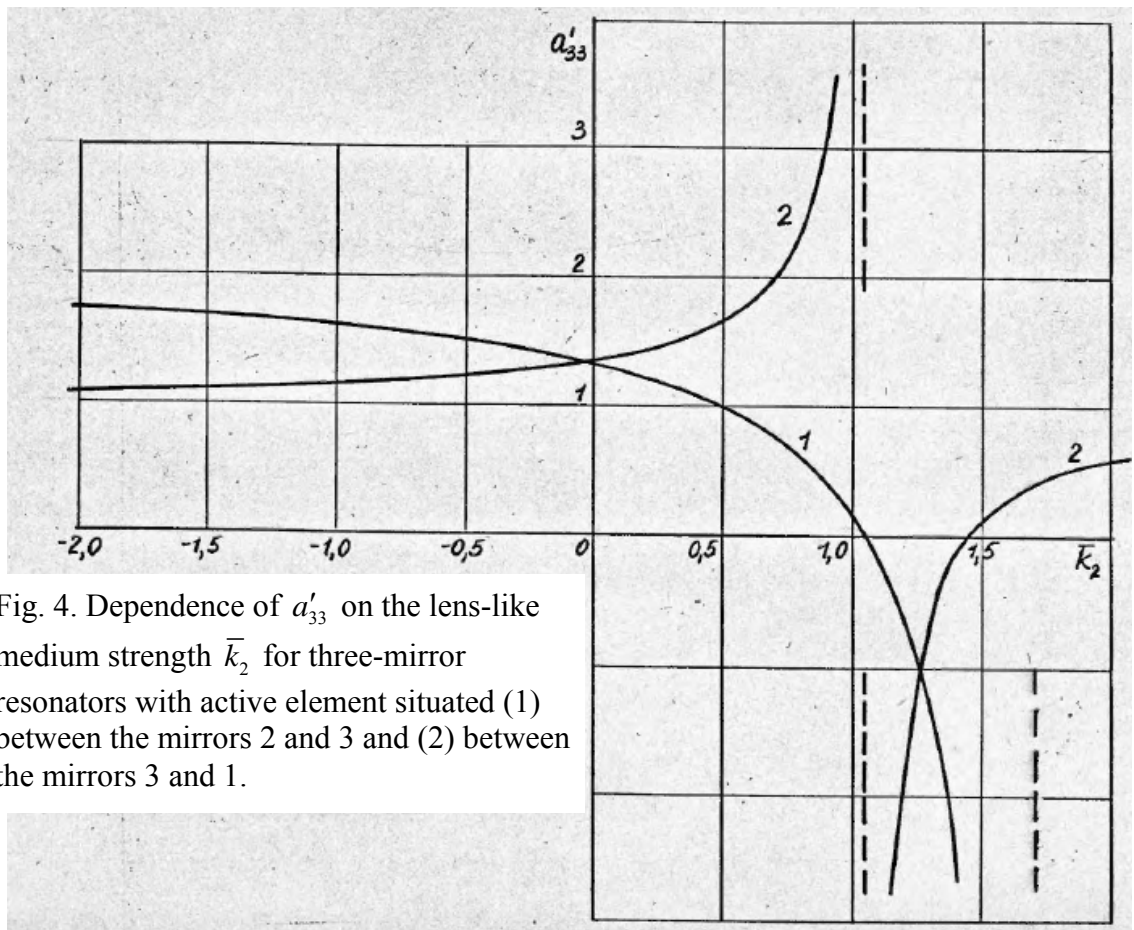

Fig. 4. Dependence of $a'_{33}$ on the lens-like medium strength $\bar{k}_2$ for three-mirror resonators with active element situated (1) between the mirrors 2 and 3 and (2) between the mirrors 3 and 1.

Noteworthy, the longitudinal distribution of the inhomogeneity, by which, in essence, the two considered cases differ, substantially influences on the output beam parameters, in contrast to the linear resonator where the mean inhomogeneity averaged over the resonator length is of the main importance [25]. Additionally, here is no strict connection between the direction of the output beam, exiting through a plane mirror, and the tilt of this mirror, which is a characteristic property of the linear resonators with transversely inhomogeneous refraction index distribution [26]. This enables, by means of the choice of parameters $R/l$ and $\eta l$, to perform regulation of almost every matrix element (except $c_{11} = d_{11} = 1/2$ in the symmetric case (b)). This suggests that ring resonators provide additional possibilities for the output beam control [26]. Generally speaking, even the simplest three-mirror resonator opens so many possibilities that their detailed analysis would only be expedient based on the limitations imposed by specific experimental conditions or by practical requirements. In this view, the results presented here can be considered as a basis for further such analysis.

In conclusion we add that, with allowance for (34), the above results supply possible means for increasing stability of the frequency characteristics by the way of proper choice of the geometric configuration and of the transverse medium inhomogeneity. Also, new approaches for the generation frequency control by means of changing the radiation spatial characteristics, and vice versa, employment of the frequency characteristics in order to control the spatial beam parameters, can be developed.

**Appendix 1. Energy method for the ring resonator calculation**

The energy method is grounded on the adiabatic invariance of the ratio between the mode energy of a conservative resonator and its eigen frequency. For linear resonators, it permits to derive, in a rather general form, some properties of the quantities analogous to the matrices (28) – (31) [21,27]. Knowledge of these properties greatly facilitates the resonator analysis; but in case of ring resonators, because of much more bulky calculations, the energy method can be especially useful.

The method is based on the fact that any resonator misalignment performs certain mechanical work on the electromagnetic field concluded in the resonator, and therefore changes its energy. According to expression (2) of Ref. [27], the energy obtained by a resonator mode when its mirrors experience deformations $\Psi_\mu(x, y)$ equals to

$$\Delta W = \sum_\mu \frac{c}{n_{0\mu}} \cos\alpha_\mu \int_{\Psi_\mu} \delta\Psi_\mu(x, y) \int_{S_\mu} (\Delta G)_{3\mu}[x, y, \Psi_\mu(x, y)] dxdy \qquad (A1)$$

(all the quantities are written in the coordinate frame of the $\mu$-th mirror, cf. Fig. 2b). Here $(\Delta G)_{3\mu}$ is the $z$-component of the electromagnetic momentum change upon reflection at the $\mu$-th mirror; expression (A1) contains integrals over the mirror surfaces $S_\mu$ as well as functional integrals over their deformations. The electromagnetic field representation in the form (1), (3) with condition (4) for $\rho_\mu = 1$ (requirement of ideal reflection is important for the conservative resonator assumption but the final results are largely independent of this assumption) gives

$$(\Delta G)_{3\mu} = \frac{n_{0\mu}^2}{2\pi\omega}\left[-2\cos\alpha_\mu |u_{\Pi\mu}|^2 + \frac{i}{2k_{0\mu}}\sin\alpha_\mu\left(u_{\Pi\mu}^* \frac{\partial u_{\Pi\mu}}{\partial x_\Pi} - u_{O\mu}^* \frac{\partial u_{O\mu}}{\partial x_O} - \text{c.c.}\right)\right] \qquad (A2)$$

Consider the misalignments of spherical mirrors of the resonator. In agreement with (11) and (5),

$$\delta\Psi_\mu(x) = dz_\mu - x\,d\overline{\beta}_\mu + d\gamma_\mu \qquad (A3)$$

and $u_\Pi$, $u_O$ are determined by expressions similar to (13). Transforming (A2) with allowance for (20) – (24) and keeping the terms quadratic in misalignments, we arrive at

$$\Delta W = \frac{c}{\omega}\sum_{\mu}\frac{n_{0\mu}}{\pi}\cos\alpha_{\mu}\left\{-\int|u_{\Pi}|^2\cos\alpha_{\mu}dxdy\left(dz_{\mu}+d\gamma_{\mu}\right)+\int|u_{\Pi}|^2 x\cos\alpha_{\mu}\,dxdy\,d\bar{\beta}_{\mu}+\right.$$

$$\left.+\sin\alpha_{\mu}\int|u_{\Pi}|^2\left[-\left(\frac{d\xi_{\Pi}}{dz_{\Pi}}\right)_{\mu}+\frac{\xi_{\Pi\mu}}{R_{\mu}\cos\alpha_{\mu}}+\bar{\beta}_{\mu}-\frac{z_{\mu}}{R_{\mu}}\tan\alpha_{\mu}\right]dxdy\,dz_{\mu}\right\}$$

where the functional integration has been reduced to the integration over the misalignment parameters $z_{\mu}$, $\gamma_{\mu}$ and $\bar{\beta}_{\mu}$.

Now, performing the integration over the mirror surfaces and taking into account that $x_{\Pi} = x\cos\alpha_{\mu} + z_{\mu}\sin\alpha_{\mu}$ we obtain

$$\Delta W = \frac{A}{\pi}\left(\frac{c}{\omega}\right)^2\sum_{\mu}\int k_{0\mu}\left\{-\cos\alpha_{\mu}\left(dz_{\mu}+d\gamma_{\mu}\right)-\left(\frac{d\xi_{\Pi}}{dz_{\Pi}}\right)_{\mu}\sin\alpha_{\mu}\,dz_{\mu}+\frac{\sin\alpha_{\mu}}{R_{\mu}\cos\alpha_{\mu}}\xi_{\Pi\mu}\,dz_{\mu}+\right.$$

$$\left.+\bar{\beta}_{\mu}\sin\alpha_{\mu}\,dz_{\mu}-z_{\mu}\sin\alpha_{\mu}\,d\bar{\beta}_{\mu}-\frac{z_{\mu}\sin^2\alpha_{\mu}}{R_{\mu}\cos\alpha_{\mu}}dz_{\mu}+\xi_{\Pi\mu}\,d\bar{\beta}_{\mu}\right\} \quad (A4)$$

where $A = \int|u_{\Pi}|^2 dx_{\Pi}dy_{\Pi}$. Further, with the help of (28) and (29), Eq. (A4) can be reduced to the form

$$\Delta W = \frac{A}{\pi}\left(\frac{c}{\omega}\right)^2\int\left\{-\sum_{\mu}k_{0\mu}\cos\alpha_{\mu}\left(dz_{\mu}+d\gamma_{\mu}\right)\right.$$

$$+\sum_{\mu,\nu}k_{0\mu}\sin\alpha_{\mu}\left(-c'_{\mu\nu}+\frac{c_{\mu\nu}}{R_{\mu}\cos\alpha_{\mu}}-\frac{\sin\alpha_{\mu}}{R_{\mu}\cos\alpha_{\mu}}\delta_{\mu\nu}\right)z_{\nu}\,dz_{\mu}$$

$$+\sum_{\mu,\nu}k_{0\mu}\sin\alpha_{\mu}\left(-a'_{\mu\nu}+\frac{a_{\mu\nu}}{R_{\mu}\cos\alpha_{\mu}}+\delta_{\mu\nu}\right)\bar{\beta}_{\nu}\,dz_{\mu}$$

$$\left.+\sum_{\mu,\nu}k_{0\mu}\left(-\sin\alpha_{\mu}\delta_{\mu\nu}+c_{\mu\nu}\right)z_{\nu}\,d\bar{\beta}_{\mu}+\sum_{\mu,\nu}k_{0\mu}a_{\mu\nu}\bar{\beta}_{\nu}\,d\bar{\beta}_{\mu}\right\}. \quad (A5)$$

Since $W$ is the function of the resonator configuration, and does not depend on the consecutive order in which the misalignment is performed, the whole integrand in (A5) is a complete differential. Consequently, if it contains a pair of terms in the form, for example, $Pdz_{\mu}+Qdz_{\nu}$, the condition

$$\frac{\partial P}{\partial z_{\nu}} = \frac{\partial Q}{\partial z_{\mu}}$$

should take place. This is quite similar to the operations with complete differential in classical thermodynamics [28]. In the integrand of (A5), the independent "state variables" are $\gamma_{\mu}$, $z_{\mu}$ and $\bar{\beta}_{\mu}$, and the problem is facilitated by the fact that corresponding multipliers (analogs of $P$ and $Q$ in the above expression) are linear functions of the state variables (the energy change $\Delta W$ is a quadratic function of the misalignment components). So the derivatives can be easily extracted from the form of (A5), and by combining all pairs of variables we arrive at

$$k_{0\nu}\left(\sin\alpha_{\nu}\cdot\delta_{\mu\nu}-c_{\nu\mu}\right)=k_{0\mu}\sin\alpha_{\mu}\left(-a'_{\mu\nu}+\frac{a_{\mu\nu}}{R_{\mu}\cos\alpha_{\mu}}+\delta_{\mu\nu}\right),$$

$$k_{0\mu}\sin\alpha_{\mu}\left(-c'_{\mu\nu}+\frac{c_{\mu\nu}}{R_{\mu}\cos\alpha_{\mu}}\right)=k_{0\nu}\sin\alpha_{\nu}\left(-c'_{\nu\mu}+\frac{c_{\nu\mu}}{R_{\nu}\cos\alpha_{\nu}}\right), \quad (A6)$$

$$k_{0\mu}a_{\mu\nu} = k_{0\nu}a_{\nu\mu}.$$

As it was in Ref. [27], analytical character of the dependence of the matrix elements (28), (29) on the misalignment parameters permits to apply the results (A6) to resonators with inhomogeneous gain or loss. Again, as was expected, the results (A6) do not depend on the assumption of ideal reflection.

Similar relations for matrices (30), (31) can be derived from (A6) by replacements $a \to -b$, $a' \to b'$, $c \to d$, $c' \to -d'$. Together with (A6) these substantially facilitate the calculations of the matrix elements for particular resonators.

Finally, we perform the integration of (A4) over the misalignment parameters with allowance for the linear dependence of $\xi_\Pi$ and $d\xi_\Pi/dz_\Pi$ on $z_\mu$ and $\bar{\beta}_\mu$. Also, with the help of (1) and (13) we can determine the electromagnetic field energy, and following to Ref. [27] we will obtain

$$\frac{\Delta\omega}{\omega} = \frac{\Delta W}{W} = \frac{1}{\Delta\Phi}\sum_\mu k_{0\mu}\left[-2\cos\alpha_\mu(z_\mu + \gamma_\mu) - \frac{z_\mu^2\sin^2\alpha_\mu}{R_\mu\cos\alpha_\mu} \right.$$
$$\left. -\left(\frac{d\xi_\Pi}{dz_\Pi}\right)_\mu z_\mu\sin\alpha_\mu + \frac{\sin\alpha_\mu}{R_\mu\cos\alpha_\mu}\xi_{\Pi\mu}z_\mu + \xi_{\Pi\mu}\bar{\beta}_\mu\right] = \frac{(\Delta\Phi)_2}{\Delta\Phi} \quad (A7)$$

where $(\Delta\Phi)_2$ is determined by equality (33). The fact that (A7) coincides with the result obtained in the usual way confirms the energy method validity for ring resonators.

**Appendix 2. Energy method generalization for a two-mirror resonator with lens-like deformation of the mirrors**

The energy-based reasoning similar to that described in Appendix 1 can be applied to more general resonator deformations provided that these do not destroy the Hermite-Gaussian eigenmode form (13). No avoid unnecessary complications, let us briefly illustrate this with the simplified example of a two-mirror (linear) resonator for which $\mu = 1, 2$ and $\alpha_\mu = 0$ (see Fig. 5). In such a resonator, $x_\Pi = x_O = x$, and the axes of the incident and reflected beams, as well as the nominal axes of the mirrors geometrically coincide and differ only by the direction. We choose the coordinate frame so that the laboratory axis $Z$ coincides with the nominal axis of the beam incident onto the right mirror 2, $Z_{\Pi 2}$ (see Fig. 5). In this case the small longitudinal misalignments $z_\mu$ do not affect the axial contour and their influence can be perfectly described via the change of the resonator length so we also take $z_\mu = 0$. However, now we admit the mirrors' deformations preserving the lens-like nature of the cavity so that, instead of (A3),

$$\delta\Psi_\mu(x) = -\frac{x^2}{2}d\left(\frac{1}{R_\mu}\right) - xd\bar{\beta}_\mu + d\gamma_\mu. \quad (A8)$$

We consider the fundamental mode for which, according to (13) and (16) – (19),

$$|u_{\Pi\mu}(x)|^2 = |u_{O\mu}(x)|^2 = \sqrt{\operatorname{Re}a_\mu}\exp\left[-\operatorname{Re}a_\mu(x - \xi_{\Pi\mu})^2\right] \quad (A9)$$

where $\operatorname{Re}a_\mu = \operatorname{Re}a_\Pi(z_\mu) = \operatorname{Re}a_O(z_\mu)$, pre-exponent factor warrants that the beam energy $\propto \int|u_\mu(x)|^2 dx$ is conserved even if the mode size changes upon the mirror deformation (A8) and, as we are restricted to the $xz$-plane, the superscripts "($x$)" are omitted. Then Eq. (A1) is transformed with account for (A8) and (A9), and the corresponding equivalent of (A4) can be found in the form

$$\Delta W = \frac{A}{\pi}\left(\frac{c}{\omega}\right)^2\sum_{\mu=1}^2 k_{0\mu}\int\left\{\frac{1}{2}\left(\xi_\mu^2 + \frac{1}{2\operatorname{Re}a_\mu}\right)d\left(\frac{1}{R_\mu}\right) + (-1)^\mu\xi_\mu d\bar{\beta}_\mu - (-1)^\mu d\gamma_\mu\right\}. \quad (A10)$$

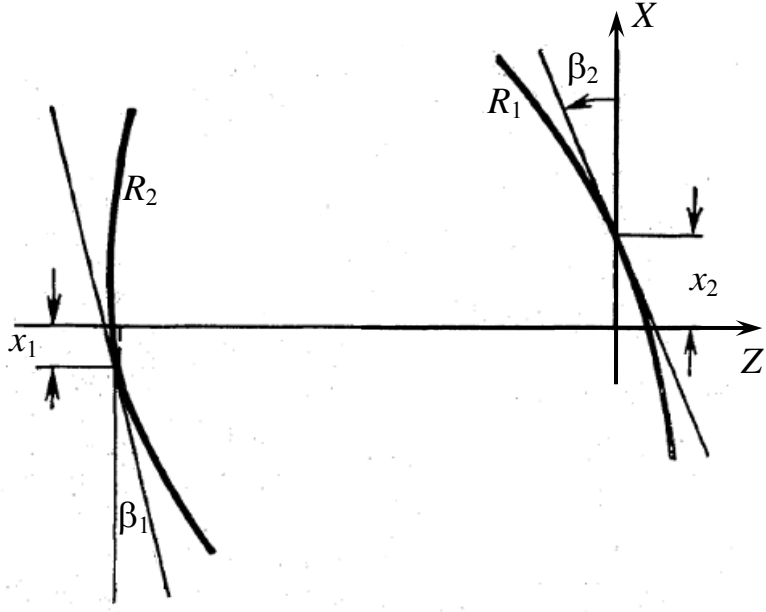

Fig. 5. Longitudinal (*XZ*) section of the two-mirror resonator (cf. Fig. 1). The mirrors with radii of curvature $R_1$ and $R_2$, are misaligned so that $\beta_1 > 0$, $\beta_2 > 0$, $x_1 < 0$, $x_2 > 0$ (cf. Eq. (11)).

In this expression, the functional integration of (A1) is reduced to integration over the parameters $1/R_\mu$, $\bar{\beta}_\mu$ and $\gamma_\mu$. Like in (A4), the integrand of (A10) is a complete differential; together with the condition that $\mathrm{Re}\, a_\mu$ does not depend on the misalignments $\bar{\beta}_\mu$ and $\gamma_\mu$, this leads to relations

$$k_0(z_1)\frac{\partial}{\partial(1/R_2)}\left(\frac{1}{\mathrm{Re}\,a_1}\right) = k_0(z_2)\frac{\partial}{\partial(1/R_1)}\left(\frac{1}{\mathrm{Re}\,a_2}\right) \tag{A11}$$

$$k_0(z_1)\frac{\partial \xi_1}{\partial \bar{\beta}_2} = -k_0(z_2)\frac{\partial \xi_2}{\partial \bar{\beta}_1} \tag{A12}$$

$$k_0(z_\mu)\frac{\partial \xi_\mu}{\partial(1/R_\nu)} = (-1)^\mu k_0(z_\nu)\xi_\nu \frac{\partial \xi_\nu}{\partial \bar{\beta}_\mu} \quad (\mu,\nu = 1,2) \tag{A13}$$

For example, (A11) gives the differential relationship between the near-mirror beam sizes and the mirrors' curvatures:

$$k_0(z_1) b_1 R_2^2 \frac{\partial b_1}{\partial R_2} = k_0(z_2) b_2 R_1^2 \frac{\partial b_2}{\partial R_1} \tag{A14}$$

where $b_\mu = (\mathrm{Re}\, a_\mu)^{-1/2}$ is the Gaussian-mode beam radius at the mirror $\mu$. One can easily persuade that (A14) agrees with the known expressions for the beam parameters in lens-like resonators (see, e.g., [24,30]).

Further, if we express the dependence $\xi_\mu(\bar{\beta}_1, \bar{\beta}_2)$ via a finite power series (polynomial), we easily find that $\xi_\mu(\bar{\beta}_1, \bar{\beta}_2)$ should be a linear homogeneous function (otherwise, the right-hand sides of Eqs. (23) contain higher powers of $\bar{\beta}_1$, $\bar{\beta}_2$ than the left-hand side, and Eqs. (23) cannot be satisfied). Of course, this property immediately follows from Eqs. (28), (30) applied to a two-mirror resonator but now it is obtained with no explicit involvement of the boundary conditions. Representing this function in the matrix form

$$\begin{pmatrix} \xi_1 \\ \xi_2 \end{pmatrix} = \begin{pmatrix} a_{11} & a_{12} \\ a_{21} & a_{22} \end{pmatrix} \begin{pmatrix} \bar{\beta}_1 \\ \bar{\beta}_2 \end{pmatrix}$$

and substituting this into Eqs. (A12) and (A13), one readily obtains the general expressions for the matrix elements:

$$\frac{1}{a_{11}} = \frac{1}{R_1} - \frac{\mathcal{C} - \mathcal{A}/R_2}{\mathcal{D} - \mathcal{B}/R_2}, \quad \frac{1}{a_{22}} = -\frac{1}{R_2} + \frac{\mathcal{C} - \mathcal{D}/R_1}{\mathcal{A} - \mathcal{B}/R_1},$$

$$\frac{1}{a_{12}} = -\frac{k_0(z_1)}{k_0(z_2)} \frac{1}{a_{21}} = \mathcal{C} - \frac{\mathcal{D}}{R_1} - \frac{\mathcal{A}}{R_2} + \frac{\mathcal{B}}{R_1 R_2} \quad (A15)$$

where the integration constants $\mathcal{A}$, $\mathcal{B}$, $\mathcal{C}$ and $\mathcal{D}$ obey the equation

$$\mathcal{AD} - \mathcal{BC} = \frac{k_0(z_1)}{k_0(z_2)}. \quad (A16)$$

The parameters $\mathcal{A}$, $\mathcal{B}$, $\mathcal{C}$ and $\mathcal{D}$ cannot be found from the "Maxwell relations" (A11) – (A13) and are determined by the special form of the lens-like optical system inside the resonator. However, even the results (A15) are useful and rather appealing. In fact, the integration constants are the elements of the ray transfer matrix [30] for the ray travelling between the mirror 1 and mirror 2 (its determinant differs from zero because of the longitudinal inhomogeneity of the axial refractive index). This result proves an important fact: if a resonator with spherical mirrors possesses an eigenmode with Gaussian intensity distribution, and upon a small misalignment the spot pattern of this mode does not change the shape but just displaces "as a whole" across the mirror surface, the optical system inside the resonator is lens-like.

For the same reason as was specified beneath (A6), formulas (A10) – (A13) and (A15), (A16) of this Appendix are applicable for complex $\xi_\mu$ in spite of the assumption that the resonator is conservative. Therefore, the interrelations between the parameters characterizing the resonator sensitivities to various sorts of deformations, (A11) – (A13), can be used for analysis of resonators with complex lens-like inhomogeneity [24].

**Appendix 3. Symmetry effects in multi-mirror resonators**

Calculations of multi-mirror resonators differ from the analysis of linear resonators by substantially larger massive of computations. It can be essentially reduced if the resonator configuration possesses symmetry, i.e. if the resonator transforms into itself upon certain mappings of the space. To this end, we consider some elementary consequences of the symmetry transformations for plane ring resonators.

There are two possible types of geometric transformations of such resonators: rotations about a certain point and reflections with respect to certain planes (symmetry axes). Since upon any reflection, the direction of a wave, propagating along the resonator, reverses, every reflection, even if it preserves the resonator configuration, actually generates a new object that differs from the initial one by the round-trip direction. To avoid the difficulties connected with this sort of symmetry, we consider the two "copies" of the resonator differing by the running wave directions as a single object.

The analysis of the spatial characteristics of ring resonators usually ends by the values of certain parameters of the resonator beam at the mirrors. It is suitable to present each set of such values in forms of 2N- dimensional vectors (N is the number of mirrors), for example:

$$(\vec{a}_\Pi, \vec{a}_O) = (a_{\Pi 1} ... a_{\Pi N}, a_{O1} ... a_{ON}),$$
$$(\vec{\xi}_\Pi, \vec{\xi}_O) = (\xi_{\Pi 1} ... \xi_{\Pi N}, \xi_{O1} ... \xi_{ON}), \quad (A17)$$

$$\left\{\left(\overrightarrow{\frac{d\xi_\Pi}{dz_\Pi}}\right),\left(\overrightarrow{\frac{d\xi_O}{dz_O}}\right)\right\} = \left\{\left(\frac{d\xi_\Pi}{dz_\Pi}\right)_1 \cdots \left(\frac{d\xi_\Pi}{dz_\Pi}\right)_N, \left(\frac{d\xi_O}{dz_O}\right)_1 \cdots \left(\frac{d\xi_O}{dz_O}\right)_N\right\}.$$

Similar representations are possible for the shifts of the output beams' axes $x_{C0}$, $y_{C0}$ (27), tilts of these axes, etc. The misalignment is defined, correspondingly, via 2$N$-vectors $(\vec{\beta}, \vec{\beta})$, $(\vec{z}, \vec{z})$ that are formed by doubling the sets of the misalignment components $\vec{\beta} = (\bar{\beta}_1 \ldots \bar{\beta}_N)$ and $\vec{z} = (z_1 \ldots z_N)$. Such a representation permits in a unified way include both copies of the resonator with opposite round-trip directions. The linear relation of the quantities $\xi_\mu$, $(d\xi/dz)_\mu$ with the misalignment parameters is expressed by block matrices of the following form:

$$\begin{pmatrix} \vec{\xi}_\Pi \\ \vec{\xi}_O \end{pmatrix} = \begin{pmatrix} (a) & 0 \\ 0 & (b) \end{pmatrix} \begin{pmatrix} \vec{\beta} \\ \vec{\beta} \end{pmatrix} + \begin{pmatrix} (c) & 0 \\ 0 & (d) \end{pmatrix} \begin{pmatrix} \vec{z} \\ \vec{z} \end{pmatrix},$$

$$\begin{pmatrix} \left(\overrightarrow{\dfrac{d\xi_\Pi}{dz_\Pi}}\right) \\ \left(\overrightarrow{\dfrac{d\xi_O}{dz_O}}\right) \end{pmatrix} = \begin{pmatrix} (a') & 0 \\ 0 & (b') \end{pmatrix} \begin{pmatrix} \vec{\beta} \\ \vec{\beta} \end{pmatrix} + \begin{pmatrix} (c') & 0 \\ 0 & (d') \end{pmatrix} \begin{pmatrix} \vec{z} \\ \vec{z} \end{pmatrix} \qquad (A18)$$

where 2$N$-vectors (A17) and vectors of misalignment are represented by columns, zeros stand for $N \times N$ zero blocks and the non-zero blocks are the matrices (28) – (31). The similar relation with misalignments exists for the quantities $x_{C0}$, $y_{C0}$, $(dx_{C0}/dz_C)$, $(dy_{C0}/dz_C)$ but because their geometrical nature is identical to that of $\xi$, $(d\xi/dz)$, it is sufficient to demonstrate the effects of symmetry on an example of matrices (A18).

We thus have arrived at considering the objects in 2$N$-dimensional space: vectors (A17), matrices (A18) (some scalars are also of interest, for example, $\Delta\Phi$) which should obey certain symmetry-induced regularities ensuing from their geometric nature.

The symmetry group of a plane figure consists of three classes whose generating elements are the unit element $E$, rotation $C_n$ through an angle $2\pi/n$ ($n \leq N$) and reflection $\sigma$ with respect to an axis crossing the rotation axis [29]. In order to determine the limitations imposed upon matrices (A18) by the symmetry, we subject both parts of any equality (A18) to a symmetry transformation, for example, $C_n$. As the resonator, after this action, transforms into itself (the structure remains unchanged),

$$C_n \begin{pmatrix} \vec{\xi}_\Pi \\ \vec{\xi}_O \end{pmatrix} = \begin{pmatrix} \vec{\xi}_\Pi \\ \vec{\xi}_O \end{pmatrix}, \quad C_n \begin{pmatrix} \vec{\beta} \\ \vec{\beta} \end{pmatrix} = \begin{pmatrix} \vec{\beta} \\ \vec{\beta} \end{pmatrix}.$$

Hence,

$$C_n \begin{pmatrix} (a) & 0 \\ 0 & (b) \end{pmatrix} C_n^{-1} = \begin{pmatrix} (a) & 0 \\ 0 & (b) \end{pmatrix}, \qquad (A19)$$

which means the invariance of the matrix with respect to transformation $C_n$. Similar reasoning shows that all matrices (A18) are invariant with respect to the rotations. When considering the reflections one should notice that the angular quantities $\bar{\beta}_\mu$, $(d\xi/dz)_\mu$ change their signs whereas the linear quantities $z_\mu$, $\xi_\mu$ do not change. This leads to requirement for matrices $(a')$, $(b')$, $(c)$, $(d)$ to be invariant and for matrices $(a)$, $(b)$, $(c')$, $(d')$ to inverse the sign upon reflections. These requirements induce certain relations between the matrix elements that enable to reduce the number of elements to be determined independently.

Within the above-specified 2N-dimensional representation, the symmetry elements are also represented by block matrices, in each row of which all elements are zeros except one equal to the unity. Blocks of the matrices $C_n$ are placed along the main diagonal and are formed from the blocks of the unit matrix by means of the circular permutation of rows (or groups of adjacent rows) of the unit matrix (the latter case corresponds to rotations combining non-neighbor mirrors, when the order of the symmetry axis is lower than the number of mirrors). The non-zero blocks of matrices $\sigma$ are situated in the upper right and lower left corners while the blocks situated on the main diagonal consist of zeros. The non-zero blocks are formed from the unit-matrix blocks by means of mutual permutations of those rows that correspond to the mirrors which are mutually replaced upon reflections.

The principle of the symmetry elements' representation can be easily understood from the explicit expressions for the elements of the group of maximal symmetry of a three-mirror resonator (the group of equilateral triangle, see Fig. 2a). It includes, except the unity element, rotations $C_1$ by 120° and $C_2$ by 240° as well as reflections in the three planes crossing the vertices and middle points of the opposite sides of the triangle.

$$E = \begin{pmatrix} \begin{matrix} 1 & 0 & 0 \\ 0 & 1 & 0 \\ 0 & 0 & 1 \end{matrix} & (0) \\ (0) & \begin{matrix} 1 & 0 & 0 \\ 0 & 1 & 0 \\ 0 & 0 & 1 \end{matrix} \end{pmatrix}, \quad C_1 = \begin{pmatrix} \begin{matrix} 0 & 1 & 0 \\ 0 & 0 & 1 \\ 1 & 0 & 0 \end{matrix} & (0) \\ (0) & \begin{matrix} 0 & 1 & 0 \\ 0 & 0 & 1 \\ 1 & 0 & 0 \end{matrix} \end{pmatrix}, \quad C_2 = \begin{pmatrix} \begin{matrix} 0 & 0 & 1 \\ 1 & 0 & 0 \\ 0 & 1 & 0 \end{matrix} & (0) \\ (0) & \begin{matrix} 0 & 0 & 1 \\ 1 & 0 & 0 \\ 0 & 1 & 0 \end{matrix} \end{pmatrix},$$

(A20)

$$\sigma_1 = \begin{pmatrix} (0) & \begin{matrix} 1 & 0 & 0 \\ 0 & 0 & 1 \\ 0 & 1 & 0 \end{matrix} \\ \begin{matrix} 1 & 0 & 0 \\ 0 & 0 & 1 \\ 0 & 1 & 0 \end{matrix} & (0) \end{pmatrix}, \quad \sigma_2 = \begin{pmatrix} (0) & \begin{matrix} 0 & 0 & 1 \\ 0 & 1 & 0 \\ 1 & 0 & 0 \end{matrix} \\ \begin{matrix} 0 & 0 & 1 \\ 0 & 1 & 0 \\ 1 & 0 & 0 \end{matrix} & (0) \end{pmatrix}, \quad \sigma_3 = \begin{pmatrix} (0) & \begin{matrix} 0 & 1 & 0 \\ 1 & 0 & 0 \\ 0 & 0 & 1 \end{matrix} \\ \begin{matrix} 0 & 1 & 0 \\ 1 & 0 & 0 \\ 0 & 0 & 1 \end{matrix} & (0) \end{pmatrix}$$

Now consider which limitations the symmetry imposes on the elements of (A18) for the three-mirror resonator, e.g., on the matrix $\begin{pmatrix} (a') & (0) \\ (0) & (b') \end{pmatrix}$. The invariance with respect to rotations $C_1$ and $C_2 = C_1^{-1}$ (A19) gives

$$a'_{11} = a'_{22} = a'_{33}, \quad a'_{12} = a'_{23} = a'_{31}, \quad a'_{21} = a'_{13} = a'_{32}. \tag{A21}$$

Obviously, similar relations hold for all other blocks of matrices entering (A18). Relations of the type (A21) reduce the total number of independent elements of each 3-row matrix to three.

Then, substituting the matrix $\sigma_1$ into the equation analogous to (A19) we obtain

$$\begin{aligned} a'_{11} &= b'_{11}, & a'_{12} &= b'_{13}, & a'_{13} &= b'_{12}; \\ a'_{21} &= b'_{31}, & a'_{22} &= b'_{33}, & a'_{23} &= b'_{32}; \\ a'_{31} &= b'_{21}, & a'_{32} &= b'_{23}, & a'_{33} &= b'_{22}; \end{aligned} \tag{A22}$$

i.e. the symmetry with respect to reflections generates links between the matrices describing the forward and backward wave characteristics. The similar relations connect the elements of matrices $(c)$ and $(d)$, $(a)$ and $-(b)$, $(c')$ and $-(d')$.

The cases of (A21) and (A22) exhaust the set of variants of incomplete symmetry for the three-mirror resonator. Any other combination of the symmetry elements is equivalent to the full symmetry including all the elements of the group (A20). Combining (A21) and (A22) we deduce that in case of complete symmetry

$$a'_{11} = a'_{22} = a'_{33} = b'_{11} = b'_{22} = b'_{33},$$
$$a'_{12} = a'_{23} = a'_{31} = b'_{13} = b'_{32} = b'_{21}, \quad (A23)$$
$$a'_{21} = a'_{13} = a'_{32} = b'_{31} = b'_{12} = b'_{23},$$

and among all elements of the matrices $(a')$, $(b')$ only three are independent. Similar equations, with account for the sign reversal in relations connecting $(a)$ and $(b)$, $(c')$ and $(d')$, take place for other matrices of (A18).

The results of this Appendix are rather elementary but not trivial, and their derivation without the symmetry analysis would be difficult. Their value is especially remarkable because they are based only on the linear relation between the beam parameters and misalignments, and are thus free from many strong limitations of the accepted model. One could hope that inevitable deviations from the lens-like approximation will affect the relations (A21) – (A23) to a significantly less degree than the particular values of the matrix elements. However, even within the frame of the lens-like model, they can be suitable for consideration of more complicated situations. In particular, equalities of the same type are valid for matrices that describe the behavior of the output beams' axes, $x_{C0}$, in case of complex lens-like inhomogeneity ($k''_2 \neq 0$).

**Appendix 4. Relations for the empty ring resonator**

In the case of empty resonator, the medium is homogeneous in all arms. Application of Eq. (17) to each arm gives

$$\xi_{\Pi\nu} = \xi_{O\mu} + \left(\frac{d\xi_O}{dx_O}\right)_\mu l, \quad \left(\frac{d\xi_\Pi}{dx_\Pi}\right)_\nu = \left(\frac{d\xi_O}{dx_O}\right)_\mu \quad (A24)$$

where $\mu$ and $\nu$ accept the values 1, 2, 3, and $\nu$ follows $\mu$ in the order of circular permutation. From (A24), in addition to (36) – (41), the equalities are obtained

$$a'_{1\nu} = b'_{3\nu}, \quad a'_{2\nu} = b'_{1\nu}, \quad a'_{3\nu} = b'_{2\nu};$$
$$c'_{1\nu} = d'_{3\nu}, \quad c'_{2\nu} = d'_{1\nu}, \quad c'_{3\nu} = d'_{2\nu}. \quad (A25)$$

Besides, due to the resonator symmetry with respect to the plane crossing the center of the spherical mirror (the vertical attitude of the triangle, Fig. 2), the relations of the type (A22) hold. Combining them with equalities (36) – (41) and (A24), (A25), the general form of the sought matrices is derived practically without algebra. The four residual quantities can be easily determined from Eqs. (35) and (A24), if we assign zero values to all the misalignment parameters except that one whose influence is characterized by the sought matrix element.

For the case (b) of Sec. 4, in (A24) one pair of equations changes

$$\xi_{\Pi 3} = \xi_{O2} \cos\eta l + \frac{1}{\eta}\left(\frac{d\xi_O}{dx_O}\right)_2 \sin\eta l, \quad \left(\frac{d\xi_\Pi}{dx_\Pi}\right)_3 = -\eta\xi_{O2}\sin\eta l + \left(\frac{d\xi_O}{dx_O}\right)_2 \cos\eta l. \quad (A26)$$

The two first columns of (A25) remain valid as well as the symmetry consequences (A22).

The case (c) differs from (b) only by that in Eqs. (A24) – (A26) the subscripts are exchanged 3→2, 2→1, and the symmetry relations (A22) do not hold.

**References.**

1. Laser metrological systems / Ed. D.P. Lukyanov. – Moscow: Radio i sviaz', 1981. – 456 p. (in Russian)
2. Ischenko Ye.F. Open optical resonators. – Moscow: Sov. radio, 1980. – 208 p. (in Russian)
3. Boytsov V.F. Influence of the active medium transverse gain distribution on the diffraction frequency splitting of counter-propagating waves in ring lasers. – Opt. i spektrosk. – 1976. – V. 41, No 5. – P. 864–869. (in Russian)
4. Boytsov V.F. Properties of the ring optical resonators with spatially-inhomogeneous medium. – Izv. vuzov. Radiofizika. – 1978. – V. 21, No 5. – P. 682–689. (in Russian)
5. Boytsov V.F. Threshold amplification and frequencies of a ring near-confocal optical resonator with spatially-inhomogeneous medium. – Opt. i spektrosk. – 1979. – V. 47, No 1. – P. 184–186. (in Russian)
6. Boytsov V.F. On the influence of the spatial inhomogeneity of the amplification coefficient and losses on the non-reciprocity of counter-propagating waves in a ring gas laser with plane mirrors. – Opt. i spektrosk. – 1980. – V. 48, No 3. – P. 611–613. (in Russian)
7. Ischenko Ye.F., Ramazanova G.S. Non-reciprocity of counter-propagating waves in a ring optical resonator containing the medium with square transverse inhomogeneity. Proc. of the Moscow Energy Institute. – 1978. – No 350. – P. 45–49. (in Russian)
8. Ischenko Ye.F., Reshetin Ye.F. Calculation of the frequency splitting in active ring resonator. – Opt. i spektrosk. – 1981. – V. 50, No 6. – P. 1062–1066. (in Russian)
9. Boytsov V.F. On the theory of the frequency non-reciprocity of the counter-propagating waves in the ring laser with plane mirrors, diaphragm and spatially-inhomogeneous medium. – Opt. i spektrosk. – 1978. – V. 45, No 2. – P. 396–387. (in Russian)
10. Boytsov V.F. Diaphragmed ring optical resonator with plane mirrors and spatially-inhomogeneous medium. – Opt. i spektrosk. – 1978. – V. 45, No 1. – P. 118–126. (in Russian)
11. Boytsov V.F. On the stability of the counter-propagating waves in a ring optical resonator with diaphragmed spherical mirror and spatially-inhomogeneous medium. – Opt. i spektrosk. – 1979. – V. 46, No 1. – P. 202–204. (in Russian)
12. Boytsov V.F., Slyusarev S.G. Ring resonator with the medium inhomogeneous in longitudinal and transverse directions and divided by a diaphragm. – Vestnik LGU. – 1981. – No 4. – P. 108–110. (in Russian)
13. Ischenko Ye.F., Reshetin Ye.F. Misalignment sensitivity of the ring optical resonator with a focusing element. – Opt. i spektrosk. – 1979. – V. 46, No 2. – P. 366–372. (in Russian)
14. Boytsov V.F., Vladimirov A.G. Properties of a ring optical resonator with the misaligned spatially-inhomogeneous medium. – Opt. i spektrosk. – 1981. – V. 51, No 4. – P. 708–713. (in Russian)
15. Boytsov V.F., Vladimirov A.G. Axial contour variation of a ring resonator upon the mirrors' misalignment. – Opt. i spektrosk. – 1982. – V. 52, No 4. – P. 724–725. (in Russian)
16. Boytsov V.F., Vladimirov A.G. Ring optical resonator with a misaligned medium. – Vestnik LGU, 1981. – 15 p. Deposited VINITI No 4013-81 Деп. (in Russian)
17. Rabinovich E.M., Melnikov L.A., Tuchin V.V. Longitudinal modes of a resonator with inhomogeneous filling. – Radiotehn. i electron. – 1979. – V. 24, No 2. – P. 328–333. (in Russian)
18. Ernst G.J., Witteman W.J. Mode structure of active resonators. – IEEE Journ. Quant. Electron., 1973, v. QE-9, No 9, p. 911–918.
19. Kravtsov Yu.A. Complex rays and complex caustics. – Izv. vuzov. Radiofizika. – 1967. – V. 10, No 9–10. – P. 1283–1304. (in Russian)

20. Louisell W.H. From Maxwell to optical resonators. // Laser Indus. Fusion and X-Ray Laser Stud. – Reading, Mass. e.a., 1976. – P. 369-485.
21. Bekshaev A.Ya. Properties of optical resonators with aberrations. PhD thesis (diss. kand. fiz.-mat. nauk). Odessa State University, Odessa, 1984. – 164 p. (in Russian)
22. Bekshaev A.Ya., Optical beam transformation on incidence on a nonplanar interface. Optics and Spectroscopy, 1984, Vol. 57, No 6, pp. 652–654.
23. Baz' A.I., Zel'dovich Y.B., Perelomov A.M. Scattering, reactions and decay in nonrelativistic quantum mechanics. – Moscow: Nauka, 1971. – 544 p. (in Russian)
24. Bekshaev A.Ya., Grimblatov V.M, Misaligned optical resonator with a lens-like medium. Soviet J. of Quantum Electr., 1980, Vol. 10, No 6, pp. 672–677.
25. Bekshaev A.Ya., Grimblatov V.M., An optical resonator with a longitudinally nonuniform lens-like medium. Journal of Applied Spectroscopy, 1983, Vol. 39, No 5, pp. 1267–1272.
26. Bekshaev A.Ya., Grimblatov V.M., Directivity of radiation leaving a misaligned resonator with a lens-like medium. Journal of Applied Spectroscopy, 1983, Vol. 39, No 1, pp. 831–834.
27. Bekshaev A.Ya., Grimblatov V.M. Energy method of analysis of optical resonators with mirror deformations. Optics and Spectroscopy, 1985, Vol. 58, No 5, pp. 707–709.
28. Sychev V.V. Differential equations of thermodynamics. – Moscow: Vysshaya shkola, 1991. – 224 p. (in Russian)
29. Zheludev I.S. Symmetry and its applications. – Moscow: Atomizdat, 1976. – 288p. (in Russian)
30. Kogelnik H., Li T. Laser beams and resonators // Applied optics. – 1966. – V. 5. – №. 10. – P. 1550-1567.